\documentstyle[aps,eqsecnum,preprint,floats,epsf,epsfig]{revtex}
\begin{document}
\def\be{\begin{eqnarray}}
\def\en{\end{eqnarray}}
\def\non{\nonumber}
\def\la{\langle}
\def\ra{\rangle}
\def\nc{N_c^{\rm eff}}
\def\vp{\varepsilon}
\def\a{{\cal A}}
\def\B{{\cal B}}
\def\c{{\cal C}}
\def\d{{\cal D}}
\def\e{{\cal E}}
\def\p{{\cal P}}
\def\t{{\cal T}}
\def\up{\uparrow}
\def\dw{\downarrow}
\def\vma{{_{V-A}}}
\def\vpa{{_{V+A}}}
\def\smp{{_{S-P}}}
\def\spp{{_{S+P}}}
\def\J{{J/\psi}}
\def\ov{\overline}
\def\Lqcd{{\Lambda_{\rm QCD}}}
\def\pr{{\sl Phys. Rev.}~}
\def\prl{{\sl Phys. Rev. Lett.}~}
\def\pl{{\sl Phys. Lett.}~}
\def\np{{\sl Nucl. Phys.}~}
\def\zp{{\sl Z. Phys.}~}
\def\lsim{ {\ \lower-1.2pt\vbox{\hbox{\rlap{$<$}\lower5pt\vbox{\hbox{$\sim$}
}}}\ } }
\def\gsim{ {\ \lower-1.2pt\vbox{\hbox{\rlap{$>$}\lower5pt\vbox{\hbox{$\sim$}
}}}\ } }

\font\el=cmbx10 scaled \magstep2{\obeylines\hfill January, 2003}

\vskip 1.5 cm

\centerline{\large\bf Hadronic Charmed Meson Decays}
\centerline{\large\bf Involving Axial Vector Mesons}
\bigskip
\centerline{\bf Hai-Yang Cheng}
\medskip
\centerline{Institute of Physics, Academia Sinica}
\centerline{Taipei, Taiwan 115, Republic of China}
\medskip

\bigskip
\bigskip
\centerline{\bf Abstract}
\bigskip
{\small Cabibbo-allowed charmed meson decays into a pseudoscalar
meson and an axial-vector meson are studied. The charm to
axial-vector meson transition form factors are evaluated in the
Isgur-Scora-Grinstein-Wise quark model. The dipole momentum
dependence of the $D\to K$ transition form factor and the presence
of a sizable long-distance $W$-exchange are the two key
ingredients for understanding the data of $D\to \ov Ka_1$.  The
$K_{1A}-K_{1B}$ mixing angle of the strange axial-vector mesons is
found to be $\approx \pm37^\circ$ or $\pm58^\circ$ from $\tau\to
K_1\nu_\tau$ decays. The study of $D\to
K_1(1270)\pi,~K_1(1400)\pi$ decays excludes the positive
mixing-angle solutions. It is pointed out that an observation of
the decay $D^0\to K_1^-(1400)\pi^+$  at the level of $5\times
10^{-4}$ will rule out $\theta\approx -37^\circ$ and favor the
solution $\theta\approx -58^\circ$. Though the decays $D^0\to \ov
K_1^0\pi^0$ are color suppressed, they are comparable to and even
larger than the color-allowed counterparts: $\ov
K_1^0(1270)\pi^0\sim K_1^-(1270)\pi^+$ and $\ov K_1^0(1400)\pi^0>
K_1^-(1400)\pi^+$. The finite width effect of the axial-vector
resonance is examined. It becomes important for $a_1(1260)$ in
particular when its width is near 600 MeV.

}

\pagebreak

\section{Introduction}

Two-body hadronic $D$ decays containing an axial-vector meson in
the final state have been studied in
\cite{Kamal,XYPham,TNPham,Kamal94,Katoch,Lipkin}. There are two
different types of axial vector mesons: $^3P_1$ and $^1P_1$, which
carry the quantum numbers $J^{\rm PC}=1^{++}$ and $1^{+-}$,
respectively. The non-strange axial vector mesons, for example,
$a_1(1260)$ and $b_1(1235)$ which correspond to $^3P_1$ and
$^1P_1$, respectively, cannot have mixing because of the opposite
$C$-parities. On the contrary, the strange partners of $a_1(1260)$
and $b_1(1235)$, namely, $K_{1A}$ and $K_{1B}$, respectively, are
not mass eigenstates and they are mixed together due to the
strange and non-strange light quark mass difference.

It has been noticed for a long time that the predicted $D^0\to
K^-a_1^+$ and $D^0\to\ov K^0 a_1^+$ rates are too small by a
factor of 5-6 and 2, respectively, when compared with experiment
\cite{Kamal,XYPham,TNPham,Kamal94,Katoch}. Interestingly, the
predicted $D^0\to K_1^-(1270)\pi^+$ and $D^+\to \ov
K_1^0(1400)\pi^+$ are also too small by roughly a factor of 5 and
2, respectively, compared to the data \cite{Katoch}. One argument
is that the factorization approach may be only suitable for
energetic two-body decays; for $D\to \ov Ka_1(1260)$ with very
little energy release, the approximation is questionable
\cite{XYPham}. Since $a_1(1260)$ is a broad resonance which will
increase the phase space available, it is thus expected that the
threshold suppression can be obviated. However, a detailed study
of the $a_1$ mass smearing effect does not show the desired
enhancement \cite{Kamal,Katoch}. Therefore, $D^0\to K^-a_1^+$ and
$D^0\to\ov K^0 a_1^+$ remain a problem. Compared to the $\rho$
production, we see experimentally that $\B(D^+\to \ov
K^0a_1^+)\gsim \B(D^+\to \ov K^0\rho^+)$ and $\B(D^0\to
K^-a_1^+)\lsim \B(D^0\to K^-\rho^+)$  \cite{PDG}. Although the
phase space for $\ov K a_1(1260)$ is largely suppressed relative
to that for $\ov K\rho$, the large $a_1(1260)$ production
comparable to $\rho$ is quite interesting. It is important to
understand these features.

The purpose of this work is to reexamine the axial-vector meson
production in the charmed meson decays and to resolve the
aforementioned long-standing problems.

The study of charm decays into an axial-vector meson and a
pseudoscalar meson will require the knowledge of form factors and
decay constants. In the early study of \cite{Katoch}, the charm to
axial vector meson transition form factors are calculated using
the ISGW (Isgur-Scora-Grinstein-Wise) model \cite{ISGW}. However,
some of the form factors get substantial modifications in the
improved version of the ISGW model, the so-called ISGW2 model
\cite{ISGW2}. For example, the relevant $D\to a_1(1260)$ and $D\to
K_{1A}$ transition form factors can be different by a factor of 3
in the ISGW and ISGW2 models.  In the present paper we will use
the ISGW2 model to compute the charm to axial-vector meson
transition form factors, and we find that $D\to\ov Ka_1(1260)$
decays provide a nice probe of the momentum dependence of the
$D\to K$ transition form factor at large $q^2$.

It is known from the data analysis based on the model-independent
diagrammatic approach \cite{CC86,Rosner} that weak annihilation
($W$-exchange or $W$-annihilation) in charm decays is quite
sizable as it can receive large contributions from final-state
interactions via quark rescattering. We shall show that the
$W$-exchange contribution is one of the key ingredients for
understanding the data.

The paper is organized as follows. In Sec. II we will discuss the
decay constants and form factors relevant for our purposes. The
$D\to AP$ decays are then discussed in detail in Sec. III. Sec. IV
gives our conclusions. An Appendix is devoted to a  sketch of the
ISGW model for the $D\to A$ transition form factor calculations.

\section{Decay constants and form factors}
In the present work we consider the isovector non-strange axial
vector mesons $a_1(1260)$ and $b_1(1235)$ and the isodoublet
strange ones $K_1(1270)$ and $K_1(1400)$. Their masses and widths
are summarized in Table I. The axial vector mesons $a_1(1260)$ and
$b_1(1235)$ have the quantum numbers $^3P_1$ and $^1P_1$,
respectively. They cannot have mixing because of the opposite
$C$-parities. However, $K_1(1270)$ and $K_1(1400)$ are a mixture
of $^3P_1$ and $^1P_1$ states owing to the mass difference of the
strange and non-strange light quarks.  We write
 \be \label{mixing}
 K_1(1270) &=& K_{1A}\sin\theta+K_{1B}\cos\theta, \non \\
 K_1(1400) &=& K_{1A}\cos\theta-K_{1B}\sin\theta,
 \en
where $K_{1A}$ and $K_{1B}$ are the strange partners of
$a_1(1260)$ and $b_1(1235)$, respectively. If the mixing angle is
$45^\circ$ and $\la K\rho|K_{1B}\ra=\la K\rho|K_{1A}\ra$, one can
show that $K_1(1270)$ is allowed to decay into $K\rho$ but not
$K^*\pi$, and vice versa for $K_1(1400)$ \cite{Lipkin77}.

\begin{table}[h]
\caption{The masses and widths of the $1\,^3P_1$ and $1\,^1P_1$
axial-vector mesons quoted in [7].
 }
\begin{center}
\begin{tabular}{l c c c c  }
 & $a_1(1260)$ & $b_1(1235)$ & $K_1(1270)$ & $K_1(1400)$ \\ \hline
 mass & $1230\pm40$ MeV & $1229.5\pm3.2$ MeV & $1273\pm7$ MeV &
 $1402\pm7$ MeV \\
 width & $250-600$ MeV & $142\pm9$ MeV & $90\pm20$ MeV &
 $174\pm13$ MeV  \\
\end{tabular}
\end{center}
\end{table}

From the experimental information on masses and the partial rates
of $K_1(1270)$ and $K_1(1400)$, Suzuki found two possible
solutions with a two-fold ambiguity, $|\theta|\approx 33^\circ$
and $57^\circ$ \cite{Suzuki}. A similar constraint $35^\circ\lsim
|\theta|\lsim 55^\circ$ is obtained in \cite{Goldman} based solely
on two parameters: the mass difference of the $a_1$ and $b_1$
mesons and the ratio of the constituent quark masses.

Based on the early data from the TPC/Two-Gamma Collaboration
\cite{TPC}
 \be
 \B(\tau^-\to K_1^-(1270)\nu_\tau) &=& (4.1^{+4.1}_{-3.5}\pm1.0)\times 10^{-3},
 \non \\ \B(\tau^-\to
 K_1^-(1400)\nu_\tau) &=& (7.6^{+4.0}_{-3.3}\pm2.0)\times 10^{-3},
 \en
Suzuki has shown that the observed $K_1(1400)$ production
dominance in the $\tau$ decay favors $|\theta|\approx 33^\circ$
\cite{Suzuki}. However, the analysis by ALEPH Collaboration based
on the LEP data yields \cite{ALEPH}
 \be
 \B(\tau^-\to K_1^-(1270)\nu_\tau) &=& (4.8\pm1.1)\times 10^{-3},
 \non \\ \B(\tau^-\to K_1^-(1400)\nu_\tau) &=& (0.5\pm1.7)\times 10^{-3}.
 \en
This indicates that $K_1(1400)$ production is somewhat reduced in
comparison with that of $K_1(1270)$. Assuming the resonance
structure of $\tau^-\to K^-\pi^+\pi^-\nu_\tau$ decays being
dominated by $K_1^-(1270)$ and $K_1^-(1400)$, both OPAL
\cite{OPAL} and CLEO \cite{CLEO} have also measured the ratio of
$K_1(1270)\nu_\tau$ to $K_1(1400)\nu_\tau$ with the averaged
result \cite{PDG}
 \be
 {\Gamma(\tau\to K_1(1270)\nu_\tau)\over \Gamma(\tau\to
K_1(1270)\nu_\tau)+\Gamma(\tau\to K_1(1400)\nu_\tau)}=0.69\pm
0.15\,.
 \en
This in turn implies that
 \be \label{K1data}
 R\equiv {\B(\tau\to K_1(1270)\nu_\tau)\over \B(\tau\to
 K_1(1400)\nu_\tau)}=2.2\pm 1.2\,.
 \en
Therefore, the new data clearly show $K_1(1270)$ dominance in the
$\tau$ decay. Consequently, the previous argument of ruling out
$|\theta|\approx 57^\circ$ from $K_1(1400)$ production dominance
is thus no longer valid. This will be elaborated in more detail
shortly below.

\subsection{Decay constants}
The decay constant of the axial-vector meson is defined by
 \be
 \la 0|A_\mu|A(q,\vp)\ra=\,f_Am_A\vp_\mu.
 \en
Because of charge conjunction invariance, the decay constant of
the $^1P_1$ non-strange neutral meson $b_1(1235)$ must be zero. In
the isospin limit, the decay constant of the charged $b_1$ must
vanish, so that $f_{b_1}$ is small. As for the strange axial
vector mesons, the $^3P_1$ and $^1P_1$ states transfer under
charge conjunction as
 \be
 M_a^b(^3P_1) \to M_b^a(^3P_1), \qquad M_a^b(^1P_1) \to
 -M_b^a(^1P_1),~~~(a,b=1,2,3).
 \en
Since the weak axial-vector current transfers as $(A_\mu)_a^b\to
(A_\mu)_b^a$ under charge conjunction, it is clear that
$f_{K_{1B}}=0$ in the SU(3) limit \cite{Suzuki}.

For $a_1(1260)$ and $K_{1A}$, their decay constants can in
principle be determined from the $\tau$ decay. From the measured
$\tau\to K_1^-(1270)\nu_\tau$ from ALEPH, the decay constant of
$K_1(1270)$ is extracted to be
 \be
 f_{K_1(1270)}=175\pm 19~{\rm MeV},
 \en
where use has been made of the formula
 \be
 \Gamma(\tau\to K_1\nu_\tau)={G_F^2\over
 16\pi}|V_{us}|^2\,f_{K_1}^2{(m_\tau^2+2m_{K_1}^2)(m_\tau^2-m_{K_1}^2)^2\over
 m_\tau^3}.
 \en
To determine the decay constant of $K_1(1400)$ we note that
$f_{K_1(1400)}/f_{K_1(1270)}=\cot\theta$ in the exact SU(3) limit.
However,  the decay constant of $K_{1B}$ is non-zero beyond the
SU(3) limit. We thus follow \cite{Suzuki} to write
 \be \label{fK1theory}
 {m_{K_1(1400)}f_{K_1(1400)}\over m_{K_1(1270)}
 f_{K_1(1270)}}={\cos\theta+\delta\sin\theta\over
 \sin\theta-\delta\cos\theta},
 \en
where in the static limit of the quark model the parameter
$\delta$ has the form \cite{Suzuki}
 \be
 |\delta|={m_s-m_u\over\sqrt{2}(m_s+m_u)}\approx 0.18\,.
 \en
The magnitude of $f_{K_1(1400)}/f_{K_1(1270)}$ can be determined
from
 \be \label{fK1expt}
 \left({f_{K_1(1400)}\over f_{K_1(1270)}}\right)^2={(m_\tau^2
 +2m_{K_1(1270)}^2)(m_\tau^2-m_{K_1(1270)}^2)^2m_{K_1(1400)}^2\over
 (m_\tau^2+2m_{K_1(1400)}^2)(m_\tau^2-m_{K_1(1400)}^2)^2m_{K_1(1270)}^2}
 \,{\Gamma(\tau\to K_1(1400)\nu_\tau)\over \Gamma(\tau\to
 K_1(1270)\nu_\tau)}.
 \en

A fit of Eqs. (\ref{fK1theory}) and (\ref{fK1expt}) to the central
value of the experimental measurement of $R$, the ratio of
$K_1(1270)\nu_\tau$ to $K_1(1400)\nu_\tau$ [see Eq.
(\ref{K1data})], yields
 \be
 \theta &=& \pm 37^\circ~~{\rm for}~\delta=\mp0.18\,, \non \\
 \theta &=& \pm 58^\circ~~{\rm for}~\delta=\pm0.18\,.
 \en
Note that these solutions for the mixing angle are consistent with
the ones $|\theta|\approx 33^\circ$ and $57^\circ$ obtained in
\cite{Suzuki} based on the partial rates of $K_1$. However,
contrary to the previous claim by Suzuki, $|\theta|\approx
58^\circ$ is still a possible solution allowed by $\tau\to
K_1\nu_\tau$ decays. In the present work we will try to see if one
of the remaining two solutions will be picked up by the study of
$D\to K_1\pi$ decays.\footnote{As pointed out by Suzuki
\cite{Suzuki97}, the relation $|M(J/\psi\to K_1^0(1400)\ov
K^0)|^2=\tan^2\theta|M(J/\psi\to K_1^0(1270)\ov K^0)|^2$ will be
able to determine $\theta$ directly without referring to other
parameters. However, these decays have thus far not been
measured.}

Although the data on $\tau\to a_1(1260)\nu_\tau\to
\rho\pi\nu_\tau$ have been reported by various experiments (for a
review, see \cite{Eidelman}), the decay $\tau\to
a_1(1260)\nu_\tau$ is not shown in the Particle Data Group
\cite{PDG}. Nevertheless, an experimental value of $f_{a_1}=203\pm
18$ MeV is quoted in \cite{Bloch}. It is generally argued that
$a_1(1260)$ should have a similar decay constant as the $\rho$
meson. This is confirmed by the model calculation, see e.g.
\cite{Isgur}. For definiteness, we choose the $a_1(1260)$ decay
constant to be 205 MeV.

\subsection {Form factors}
Form factors for the $D\to P$ transition are defined by \cite{BSW}
 \be \label{m.e.}
 \la P(p)|V_\mu|D(p_D)\ra = \left(p_{D\mu}+p_\mu-{m_D^2-m_{P}^2\over q^2}\,q_ \mu\right)
F_1^{DP}(q^2)+{m_D^2-m_{P}^2\over q^2}q_\mu\,F_0^{DP}(q^2),
 \en
where $q_\mu=(p_D-p)_\mu$. One of the form factors relevant for
$D\to AP$ decays is $F_1^{DP}(q^2)$. To compute this form factor
we will use the Bauer-Stech-Wirbel (BSW) model \cite{BSW} which
adopts the pole dominance assumption for the form-factor momentum
dependence
 \be
 f(q^2)={f(0)\over (1-q^2/m_*^2)^n},
 \en
with $m_*$ being the $1^-$ ($0^-$) pole mass for $F_1$ ($F_0$).
The original BSW model assumes a monopole behavior (i.e. $n=1$)
for all the form factors. However, this is not consistent with
heavy quark symmetry scaling relations for heavy-to-light
transitions. The modified BSW model takes the BSW model results
for the form factors at zero momentum transfer but makes a
different ansatz for their $q^2$ dependence, namely, a dipole
behavior (i.e. $n=2$) is assumed for the form factors
$F_1,~V_0,~V_2,~A$, motivated by heavy quark symmetry, and a
monopole dependence for $F_0,V_1$, where the form factors $V_i$
and $A$ will be introduced shortly.

In the Isgur-Scora-Grinstein-Wise (ISGW) model \cite{ISGW,ISGW2},
the vector form factors for $D\to A$ transition are defined by
 \be \label{DAform}
 \la A(p_A,\vp)(^3P_1)|V_\mu|D(p_D)\ra &=& \ell\vp_\mu^*+c_+(\vp^*\cdot
 p_D)(p_D+p_A)_\mu+c_-(\vp^*\cdot p_D)(p_D-p_A)_\mu, \non \\
 \la A(p_A,\vp)(^1P_1)|V_\mu|D(p_D)\ra &=& r\vp_\mu^*+s_+(\vp^*\cdot
 p_D)(p_D+p_A)_\mu+s_-(\vp^*\cdot p_D)(p_D-p_A)_\mu.
 \en
The form factors $\ell$, $c_+$, $c_-$, $r$, $s_+$ and $s_-$ can be
calculated in the ISGW quark model \cite{ISGW} and its improved
version, the ISGW2 model \cite{ISGW2}. In general, the form
factors evaluated in the ISGW model are reliable only at
$q^2=q^2_m\equiv (m_D-m_A)^2$, the maximum momentum transfer. The
reason is that the form-factor $q^2$ dependence in the ISGW model
is proportional to exp[$-(q^2_m-q^2)$] (see Eq. (\ref{oldFn})) and
hence the form factor decreases exponentially as a function of
$(q^2_m-q^2)$. This has been improved in the ISGW2 model in which
the form factor has a more realistic behavior at large
$(q^2_m-q^2)$ which is expressed in terms of a certain polynomial
term (see Eq. (\ref{Fn})). In addition to the form-factor momentum
dependence, the ISGW2 model incorporates a number of improvements,
such as the constraints imposed by heavy quark symmetry, hyperfine
distortions of wave functions, etc.,$\cdots$ \cite{ISGW2}.

Note that the results for the form factor $c_+$ are quite
different in the ISGW and ISGW2 models (see Table II): $c_+$ is
positive in the former model while it becomes negative in the
latter (see the Appendix for details).

\begin{table}[ht]
\caption{The form factors at $q^2=m_K^2$ for $D\to a_1$ and $D\to
b_1$ transitions and at $q^2=m_\pi^2$ for $D\to K_{1A}$ and $D\to
K_{1B}$ transitions, where $\ell$ and $r$ are in units of GeV and
others carry units of ${\rm GeV}^{-1}$. The first entry is for the
form factors calculated in the ISGW model and the second entry is
for the ISGW2 model.
 }
\begin{center}
\begin{tabular}{l c c c c c c  }
Transition & $\ell$ & $c_+$ & $c_-$ & $r$ & $s_+$ & $s_-$ \\
\hline
 $D\to a_1$ & $-0.93$ & $0.20$ & & & & \\
 & $-1.31$ & $-0.11$ & $-0.037$ & & &  \\
 $D\to b_1$ & & & & 0.95 & 0.42 & \\
 & & & & 1.29 & 0.20 & $-0.072$ \\
 $D\to K_{1A}$ & $-0.49$ & 0.12 & & & & \\
 & $-0.78$ & $-0.13$ & $-0.030$ & & & \\
 $D\to K_{1B}$ & & & & 0.64 & 0.31 & \\
 & & & &  0.94 & 0.21 & $-0.051$ \\
\end{tabular}
\end{center}
\end{table}

In realistic calculations of decay amplitudes it is convenient to
use the dimensionless form factors defined by \cite{BSW}
 \be
 \la A(p_A,\vp)|V_\mu|D(p_D)\ra &=&
\Bigg\{(m_D+m_A) \vp^*_\mu V_1^{DA}(q^2)  - {\vp^*\cdot
p_D\over m_D+m_A}(p_D+p_A)_\mu V_2^{DA}(q^2) \non \\
&-& 2m_A {\vp^*\cdot p_D\over
q^2}(p_D-p_A)_\mu\left[V_3^{DA}(q^2)-V_0^{DA}(q^2)\right]\Bigg\},
\non \\
  \la A(p_A,\vp)|A_\mu|D(p_D)\ra &=& {2\over
  m_D+m_A}i\epsilon_{\mu\nu\rho\sigma}\vp^{*\nu}p_D^\rho p_A^\sigma
  A^{DA}(q^2),
 \en
with
 \be V_3(q^2)=\,{m_D+m_A\over 2m_A}\,V_1(q^2)-{m_D-m_A\over
2m_A}\,V_2(q^2),
 \en
and $V_3(0)=V_0(0)$. Note that only the form factor $V_0$ will
contribute to the factorizable amplitude as one can check the
matrix element $q^\mu \la A(p_A,\vp)|V_\mu|D(p_D)\ra$. The ISGW
and ISGW2 model predictions for the form factors $V_{0,1,2}$ are
exhibited in Table III.

\begin{table}[ht]
\caption{The dimensionless vector form factors $V_{0,1,2}$ at
$q^2=m_K^2$ for $D\to a_1$ and $D\to b_1$ transitions and at
$q^2=m_\pi^2$ for $D\to K_{1A}$ and $D\to K_{1B}$ transitions
calculated in the ISGW2 model. The numbers in parentheses are the
results obtained using the ISGW model.
 }
\begin{center}
\begin{tabular}{l c c c  }
Transition & $V_0$ & $V_1$ & $V_2$ \\
\hline
 $D\to a_1$ & $-0.63~(-0.22)$ & $-0.42~(-0.30)$ & $0.35~(-0.63)$ \\
 $D\to b_1$ & 0.68~(0.72) & 0.42~(0.31) & $-0.62~(-1.29)$ \\
 $D\to K_{1A}$ & $-0.37~(-0.11)$ & $-0.24~(-0.15)$ & $0.40~(-0.39)$ \\
 $D\to K_{1B}$ & 0.50~(0.45) & 0.29~(0.20) & $-0.65~(-0.99)$ \\
\end{tabular}
\end{center}
\end{table}

\section{$D\to AP$ decays}
We will study some of the Cabibbo-allowed $D\to AP$ decays ($A$:
axial-vector meson, $P$: pseudoscalar meson) within the framework
of generalized factorization in which the hadronic decay amplitude
is expressed in terms of factorizable contributions multiplied by
the {\it universal} (i.e. process independent) effective
parameters $a_i$ that are renormalization scale and scheme
independent. More precisely, the weak Hamiltonian has the form
 \be
 H_{\rm eff}={G_F\over\sqrt{2}}V_{cs}V_{ud}^*\Big[ a_1(\bar ud)
 (\bar sc)+a_2(\bar sd)(\bar uc)\Big]+h.c.,
 \en
with $(\bar q_1q_2)\equiv \bar q_1\gamma_\mu(1-\gamma_5)q_2$. For
hadronic charm decays, we shall use $a_1=1.15$ and $a_2=-0.55$\,.
The parameters $a_1$ and $a_2$ are related to the Wilson
coefficients via
 \begin{eqnarray}  \label{a12}
a_1= c_1(\mu) + c_2(\mu) \left({1\over N_c} +\chi_1(\mu)\right)\,,
\qquad \quad a_2 = c_2(\mu) + c_1(\mu)\left({1\over N_c} +
\chi_2(\mu)\right)\,,
 \end{eqnarray}
where the nonfactorizable terms $\chi_i(\mu)$ will compensate the
scale and scheme dependence of Wilson coefficients $c_i(\mu)$ to
render $a_i$ physical.

In terms of the topological amplitudes: $T$, the color-allowed
external $W$-emission tree diagram; $C$, the color-suppressed
internal $W$-emission diagram; $E$, the $W$-exchange diagram, the
Cabibbo-allowed $D\to A\pi$ $(A=K_1(1270),~K_1(1400)$) and
$D\to\ov K A$ ($A=a_1(1260),~b_1(1235)$) amplitudes have the
expressions:
 \be
 && A(D^0 \to A^-\pi^+) =T+E, \qquad A(D^0\to A^0 \pi^0)={1\over\sqrt{2}}(C'-E),
 \non \\
 && \qquad\qquad A(D^+\to A^0 \pi^+)= T+C',
 \en
and
  \be
 && A(D^0 \to K^-A^+) =T'+E, \qquad A(D^0\to \ov K^0 A^0)={1\over\sqrt{2}}(C-E),
 \non \\
 && \qquad \qquad A(D^+\to \ov K^0 A^+)= T'+C.
 \en
For $D\to AP$ and $D\to PA$ decays, one can have two different
external $W$-emission and internal $W$-emission diagrams,
depending on whether the emission particle is a scalar meson or a
pseudoscalar one. We thus denote the prime amplitudes $T'$ and
$C'$ for the case when the scalar meson is an emitted particle
\cite{ChengDSP}.

\subsection{$D\to Ka_1(1260)$ and $D\to Kb_1(1235)$}
Under the factorization approximation, the $D\to Ka_1(1260)$ and
$D\to Kb_1(1235)$ decay amplitudes read (the overall $\vp^*\cdot
p_D$ terms being dropped for simplicity)
 \be
 A(D^+\to \ov K^0a_1^+(1260)) &=&
 {G_F\over\sqrt{2}}V_{cs}V_{ud}^*\left[
 2a_1f_{a_1}m_{a_1}F_1^{DK}(m_{a_1}^2)+2a_2f_Km_{a_1}V_0^{Da_1}(m_K^2)\right],
 \non \\
 A(D^0\to K^-a_1^+(1260)) &=& {G_F\over\sqrt{2}}V_{cs}V_{ud}^*\,
 2a_1f_{a_1}m_{a_1}F_1^{DK}(m_{a_1}^2),   \non \\
 A(D^0\to \ov K^0a_1^0(1260)) &=&
 {G_F\over 2}V_{cs}V_{ud}^*\,2a_2f_Km_{a_1}V_0^{Da_1}(m_K^2),
 \en
and
  \be
 A(D^+\to \ov K^0b_1^+(1235)) &=&
 {G_F\over\sqrt{2}}V_{cs}V_{ud}^*\left[
 2a_1f_{b_1}m_{b_1}F_1^{DK}(m_{b_1}^2)+2a_2f_Km_{b_1}V_0^{Db_1}(m_K^2)\right],
 \non \\
 A(D^0\to K^-b_1^+(1235)) &=& {G_F\over\sqrt{2}}V_{cs}V_{ud}^*\,
 2a_1f_{b_1}m_{b_1}F_1^{DK}(m_{b_1}^2),   \non \\
 A(D^0\to \ov K^0b_1^0(1235)) &=&
 {G_F\over 2}V_{cs}V_{ud}^*\,2a_2f_Km_{b_1}V_0^{Db_1}(m_K^2),
 \en
where the factorizable $W$-exchange amplitude has been neglected
owing to helicity and color suppression.

As mentioned in the Introduction, the branching ratios of the
decays $D^0\to K^-a_1^+(1260)$ and $D^+\to \ov K^0a_1^+(1260)$
have been predicted to be of order 1.5\% and 3.8\%, respectively
\cite{Kamal} which are well below the measured values of $(7.2\pm
1.1)\%$ and $(8.1\pm1.7)\%$ (see Table IV). In our study, the $\ov
K^0a_1^+$ rate gets enhanced for two reasons: (i) The $q^2$
dependence of the form factor $F_1^{DK}(q^2)$ is of the dipole
rather than the monopole form in order to be consistent with heavy
quark symmetry.\footnote{If we use the Melikhov-Stech (MS) model
\cite{MS} to evaluate the $D\to K$ transition form factor, the
branching ratios will become 6.9\% and 3.3\%, respectively, for
$\ov K^0a_1^+$ and $K^-a_1^+$. This implies that the increase of
$F_1^{DK}(q^2)$ at $q^2=m_{a_1(1260)}^2$ is not fast enough in
this phenomenological model. More precisely, $F_1^{DK}(0)=0.78$
and $F_1^{DK}(m_{a_1}^2)=1.29$ in the MS model, while the
corresponding values are 0.76 and 1.75 in the improved BSW model.}
(ii) Contrary to \cite{Kamal} where the form factor $V_0^{Da_1}$
is assumed to be zero, the calculated form factor using the ISGW2
model yields a negative $V_0$ for $D\to a_1$ transition and a
positive one for $D\to b_1$. This means that the interference
between external and internal $W$-emission amplitudes is
constructive  in $D^+\to \ov K^0 a_1^+(1260)$ and destructive in
$D^+\to \ov K^0 b_1^+(1235)$. Our result for the former is
slightly larger than experiment (see Table IV). Recall that this
mode has been measured by two different groups with the branching
ratios of $(11.6\pm3.7)\%$ by E691 \cite{E691} and $(7.5\pm1.6)\%$
by Mark III \cite{Mark3}. Therefore, our result is in good
agreement with E691. In view of this, it is important to have a
refined measurement of this decay mode.

\begin{table}[ht]
\caption{Branching ratios for $D\to Ka_1(1260)$ and $D\to
Kb_1(1235)$.
 }
\begin{center}
\begin{tabular}{l c c c}
&  \multicolumn{2}{c}{Theory}  \\ \cline{2-3}
 \raisebox{1.5ex}[0cm][0cm]{Decay}
& without FSIs & with FSIs &
 \raisebox{1.5ex}[0cm][0cm]{Experiment \cite{PDG}} \\
\hline
 $D^+\to \ov K^0a_1^+(1260)$ & 12.1\% & 12.1\% & $(8.1\pm1.7)\%$  \\
 $D^0\to K^-a_1^+(1260)$ & 3.8\% & 6.2\% & $(7.2\pm 1.1)\%$ \\
 $D^0\to \ov K^0a_1^0(1260)$ & $3.3\times 10^{-4}$ & $5.6\times 10^{-4}$ & $<1.9\%$ \\
 \hline
 $D^+\to \ov K^0b_1^+(1235)$ & $1.7\times 10^{-3}$ & $1.7\times 10^{-3}$ &  \\
 $D^0\to K^-b_1^+(1235)$ & $3.7\times 10^{-6}$ & $5.9\times 10^{-6}$ & \\
 $D^0\to \ov K^0b_1^0(1235)$ & $3.9\times 10^{-4}$ & $6.7\times 10^{-4}$ &  \\
\end{tabular}
\end{center}
\end{table}

As for $D^0\to K^-a_1^+(1260)$, the dipole $q^2$ dependence of the
form factor $F_1^{DK}$ will enhance its branching ratio from 1.7\%
to 3.8\% (see the second column of Table IV). However, it is still
smaller than experiment by a factor of 2. This is ascribed to the
fact that we have so far neglected the $W$-exchange contribution.
It has been noticed that a large long-distance $W$-exchange can be
induced from final-state rescattering (see e.g. \cite{a1a2charm}).
The data analysis of Cabibbo-allowed $D\to \ov K\rho$ decays
indicates \cite{Rosner}
 \be \label{E/T}
 \left.{E\over T}\right|_{D\to \ov K\rho}\approx
 0.54\,e^{-i72^\circ}, \qquad \left.{E\over C}\right|_{D\to \ov K\rho}\approx
 1.12\,e^{i76^\circ}.
 \en
If we assume that this result holds also for $D\to \ov KA$
($A=a_1(1260),~b_1(1235)$), then the branching ratio will be
enhanced to 6.2\% as shown on the third column of Table IV. We
also see that the FSI induced $W$-exchange will increase the
branching ratio of $D^0\to\ov K^0a_1^0(1260)$ from $3.3\times
10^{-4}$ to $5.6\times 10^{-4}$.

It is interesting to notice that although the phase space for the
final state $\ov Ka_1(1260)$ is substantially suppressed relative
to $\ov K\rho$, the large $D\to K$ transition form factor at
$q^2=m_{a_1}^2$ and the negative form factor $V_0$ for $D\to a_1$
transition render $\B(D^+\to \ov K^0a_1^+)\gsim \B(D^+\to \ov
K^0\rho^+)$ and $\B(D^0\to K^-a_1^+)\lsim \B(D^0\to K^-\rho^+)$.
However, $\B(D^0\to \ov K^0a_1^0)< \B(D^0\to \ov K^0\rho^0)$.

Owing to the smallness of the $b_1$ decay constant, the decay
rates of $\ov K^0 b_1^+$ and $K^-b_1^+$ are much smaller than
their counterparts $\ov K^0a_1^+$ and $K^-a_1^+$. Nevertheless,
the neutral modes $\ov K^0b_1^0$ and $\ov K^0a_1^0$ are
comparable.

\subsection{$D\to K_1(1270)\pi$ and $D\to K_1(1400)\pi$}
The factorizable amplitudes for $D\to K_1(1270)\pi$ and $D\to
K_1(1400)\pi$ are (the overall $\vp^*\cdot p_D$ terms being
dropped for simplicity)\footnote{In \cite{Katoch}, the
color-suppressed amplitudes in $D\to \ov K_1(1270)\pi$ and $\ov
K_1(1400)\pi$ decays characterized by the parameter $a_2$ are
erroneously multiplied by an additional factor of $\sin\theta$ and
$\cos\theta$, respectively.}
 \be \label{K1pi}
 A(D^+\to \ov K^0_1(1270)\pi^+) &=&
 {G_F\over\sqrt{2}}V_{cs}V_{ud}^*\Big[2a_1m_{K_1(1270)}f_\pi(\sin\theta\,
 V_0^{DK_{1A}}(m_\pi^2)+\cos\theta\, V_0^{DK_{1B}}(m_\pi^2))  \non \\
 &+& 2a_2m_{K_1(1270)}f_{K_1(1270)}F_1^{D\pi}(m^2_{K_1(1270)})\Big], \non \\
 A(D^+\to \ov K^0_1(1400)\pi^+) &=&
 {G_F\over\sqrt{2}}V_{cs}V_{ud}^*\Big[2a_1m_{K_1(1400)}f_\pi(\cos\theta\,
 V_0^{DK_{1A}}(m_\pi^2)-\sin\theta\, V_0^{DK_{1B}}(m_\pi^2))  \non \\
 &+& 2a_2m_{K_1(1400)}f_{K_1(1400)}F_1^{D\pi}(m^2_{K_1(1400)})\Big], \non \\
  A(D^0\to K^-_1(1270)\pi^+) &=&
 {G_F\over\sqrt{2}}V_{cs}V_{ud}^*\Big[2a_1m_{K_1(1270)}f_\pi(\sin\theta\,
 V_0^{DK_{1A}}(m_\pi^2)+\cos\theta\, V_0^{DK_{1B}}(m_\pi^2))\Big], \non \\
 A(D^0\to K^-_1(1400)\pi^+) &=&
 {G_F\over\sqrt{2}}V_{cs}V_{ud}^*\Big[2a_1m_{K_1(1400)}f_\pi(\cos\theta\,
 V_0^{DK_{1A}}(m_\pi^2)-\sin\theta\, V_0^{DK_{1B}}(m_\pi^2))\Big], \non \\
 A(D^0\to \ov K^0_1(1270)\pi^0) &=&
 {G_F\over 2}V_{cs}V_{ud}^*\Big[2a_2m_{K_1(1270)}f_{K_1(1270)}F_1^{D\pi}(m^2_{K_1(1270)})\Big], \non \\
 A(D^0\to \ov K^0_1(1400)\pi^0) &=&
 {G_F\over 2}V_{cs}V_{ud}^*\Big[
 2a_2m_{K_1(1400)}f_{K_1(1400)}F_1^{D\pi}(m^2_{K_1(1400)})\Big],
 \en
where we have taken into account the $K_{1A}-K_{1B}$ mixing given
by Eq. (\ref{mixing}). As before, we have neglected the
short-distance factorizable $W$-exchange contribution.

Using the $D\to K_{1A}$ and $D\to K_{1B}$ form factors computed in
the ISGW2 model (see Table III) and $f_{K_1(1270)}=145$ MeV, the
results for the branching ratios of $D\to K_1\pi$ are depicted in
Table V for the mixing angles $|\theta|=37^\circ$ and $58^\circ$.
It is evident that the positive mixing-angle solutions
$\theta=37^\circ$ and $58^\circ$ are ruled out as the predicted
$\ov K_1^0(1270)\pi^+$ is too large while $K^-(1270)\pi^+$ is too
small compared to experiment. Note that the experimental limit on
$D^+\to\ov K^0_1(1270)\pi^+$ is measured to be 0.007 by E691
\cite{E691} and 0.011 by Mark III \cite{Mark3}. Therefore, both
negative mixing-angle solutions are allowed by experiment.
However, $D^0\to K_1^-(1400)\pi^+$ is very suppressed for
$\theta\approx -37^\circ$. Hence an observation of this mode at
the level of $5\times 10^{-4}$ will rule out $\theta\approx
-37^\circ$ and favor the other solution $\theta\approx -58^\circ$.

\begin{table}[ht]
\caption{Branching ratios of $D\to K_1(1270)\pi$ and $D\to
K_1(1400)\pi$ calculated for various $K_{1A}-K_{1B}$ mixing
angles.
 }
\begin{center}
\begin{tabular}{lcccccc}
&  \multicolumn{4}{c}{Theory}  \\ \cline{2-5}
 \raisebox{1.5ex}[0cm][0cm]{Decay}
& $-37^\circ$ & $-58^\circ$ & $37^\circ$ & $58^\circ$ &
 \raisebox{1.5ex}[0cm][0cm]{Experiment \cite{PDG}} \\
 \hline
 $D^+\to \ov K_1^0(1270)\pi^+$ & $6.4\times 10^{-3}$ & $7.8\times 10^{-3}$
 & $2.9\times 10^{-2}$ & $4.7\times 10^{-2}$ & $<7\times 10^{-3}$ \\
 $D^+\to \ov K_1^0(1400)\pi^+$ & $2.9\times 10^{-2}$ & $4.0\times 10^{-2}$
 & $6.6\times 10^{-2}$  & $6.6\times 10^{-2}$ & $(4.9\pm1.2)\%$ \\
 $D^0\to K^-_1(1270)\pi^+$ & $6.3\times 10^{-3}$ & $5.5\times 10^{-3}$
 & $4.9\times 10^{-4}$  & $4.4\times 10^{-5}$ & $(1.13\pm 0.31)\%$ \\
 $D^0\to K^-_1(1400)\pi^+$ & $3.7\times 10^{-8}$ & $4.2\times 10^{-4}$
 & $3.0\times 10^{-3}$ & $3.2\times 10^{-3}$ & $<1.2\%$ \\
 $D^0\to \ov K^0_1(1270)\pi^0$ & $8.4\times 10^{-3}$ & $8.4\times 10^{-3}$
 & $8.4\times 10^{-3}$ & $8.4\times 10^{-3}$ & $<2.0\%$ \\
 $D^0\to \ov K^0_1(1400)\pi^0$ & $5.7\times 10^{-3}$ & $5.5\times 10^{-3}$
 & $5.7\times 10^{-3}$ & $5.5\times 10^{-3}$ & $<3.7\%$ \\
\end{tabular}
\end{center}
\end{table}

Several remarks are in order. (i) For the decay constant of
$K_1(1270)$, we use the value of 145 MeV rather than 175 MeV as
inferred from the $\tau\to K_1(1270)\nu_\tau$ decay. If the latter
is used, we will have $\B(D^+\to\ov K_1^0(1270)\pi^+)=1.5\%$ and
1.7\%, respectively, for $\theta=-37^\circ$ and $-58^\circ$, which
exceed the current experimental limit. (ii) In Table V we have not
taken into account the $W$-exchange contributions. If we assume
that the $W$-exchange term relative to the amplitudes $T$ and $C$
is similar to that in $D\to \ov K^*\pi$ decays, namely
\cite{Rosner},
 \be \label{K*pi}
 \left.{E\over T}\right|_{D\to \ov K^*\pi}\approx
 0.78\,e^{i96^\circ}, \qquad \left.{E\over C}\right|_{D\to \ov K^*\pi}\approx
 0.94\,e^{i248^\circ},
 \en
the branching ratios of $\ov K_1^0(1270)\pi^0$ and $\ov
K_1^0(1400)\pi^0$ will become 2.2\% and 1.4\%, respectively. The
former slightly exceeds the current limit. Therefore, the
realistic value of $W$-exchange is smaller than that given by Eq.
(\ref{K*pi}). (iii) We see that $\ov K_1^0(1400)\pi^+$ is larger
than $\ov K_1^0(1270)\pi^+$ by one order of magnitude since the
interference between color-allowed and color-suppressed amplitudes
is constructive in the latter and destructive in the former. (iv)
Though the decays $D^0\to \ov K_1^0\pi^0$ are color suppressed,
they are comparable to and even larger than the color-allowed
counterparts: $\ov K_1^0(1270)\pi^0\sim K_1^-(1270)\pi^+$ and $\ov
K_1^0(1400)\pi^0> K_1^-(1400)\pi^+$. This can be seen from Eq.
(\ref{K1pi}) and from the fact that the form factor $V_0$ is
negative (positive) for $D\to K_{1A}$ ($D\to K_{1B}$) transition
and that $F_1^{D\pi}$ is large at $q^2=m_{K_1(1270)}^2$ or
$m_{K_1(1400)}^2$. Since the inclusion of the $W$-exchange
contribution will enhance the decay rates of $\ov
K_1^0(1270)\pi^0$ and $\ov K_1^0(1400)\pi^0$ by a factor of, say
1.5, it is conceivable that $D^0\to \ov K_1^0\pi^0$ has a
branching ratio of order $10^{-2}$. Hence, the neutral $\ov
K_1^0\pi^0$ modes should be easily accessible by experiment.

\subsection{Finite width effect}
Among the four axial-vector mesons we have studied thus far,
$a_1(1260)$ is a broad resonance with a large width ranging from
250 MeV to 600 MeV and hence it will increase the phase space
available. A running mass for the resonance has been considered in
\cite{Kamal} to take into account the smearing effect due to the
large width. However, the ansatz of a Breit-Wigner measure
$\rho(m^2)$ made in \cite{Kamal} is somewhat arbitrary.

The factorization relation
 \be \label{fact}
 \Gamma(D\to RM\to M_1M_2M)=\Gamma(D\to RM)\B(R\to M_1M_2),
 \en
which is often employed is, strictly speaking, valid only in the
narrow width approximation. For an illustration, we consider the
decay $D\to \ov K a_1(1260)\to \ov K \pi\pi\pi$. Following
\cite{Chengf0}, we compute the quantity
  \be
 \eta\equiv {\Gamma(D\to \ov K a_1(1260)\to \ov K \rho\pi \to
 \ov K  \pi\pi\pi)\over \Gamma(D\to \ov K a_1(1260))
 \B(a_1\to \rho\pi\to \pi\pi\pi)},
 \en
where we have assumed that $a_1(1260)$ decays entirely into
$\rho\pi$ \cite{PDG}. The deviation of $\eta$ from unity will give
a measure of the violation of the factorization relation. Owing to
the finite width effect, the effective decay rate of $D\to \ov K
a_1(1260)$ becomes
 \be
 \Gamma(D\to\ov Ka_1(1260))_{\rm fw}=\,\eta\, \Gamma(D\to\ov
 Ka_1(1260)).
 \en

To proceed we write the on-shell decay amplitudes as
 \be
 A(D\to \ov Ka_1)=M(D\to\ov Ka_1)(\vp^*\cdot p_D), \qquad
 A(a_1\to\rho\pi)=(gG_{\mu\nu}+hL_{\mu\nu})\vp_{a_1}^\mu\vp^{*\nu}_\rho,
 \en
where \cite{Bloch}
 \be
 G^{\mu\nu} &=& \delta^{\mu\nu}-{1\over Y}\left[ m_{a_1}^2
 p_\rho^\mu p_\rho^\nu+m_\rho^2 p_{a_1}^\mu p_{a_1}^\nu+p_{a_1}\cdot
 p_\rho(p_{a_1}^\mu p_{\rho}^\nu+p_{\rho}^\mu p_{a_1}^\nu)\right], \non \\
 L^{\mu\nu} &=& {p_{a_1}\cdot p_\rho\over
 Y}\left(p_{a_1}^\mu+p_{\rho}^\mu {m_{a_1}^2\over p_{a_1}\cdot
 p_\rho}\right)\left(p_{\rho}^\nu+p_{a_1}^\nu{m_\rho^2\over p_{a_1}\cdot
 p_\rho}\right),
 \en
and $Y=(p_{a_1}\cdot p_\rho)^2-m_{a_1}^2m_\rho^2$. The two-body
decay rates then read
 \be \label{decayrate}
 \Gamma(D\to \ov Ka_1) &=& {p^3\over 8\pi m_D^2}|M(D\to \ov Ka_1)|^2,
 \non \\  \Gamma(a_1\to \rho\pi) &=& {p'\over 12\pi
 m_{a_1}^2}|M(a_1\to\rho\pi)|^2,
 \en
where \cite{Bloch}
 \be
 |M(a_1\to\rho\pi)|^2=\left(2|g|^2+{m_{a_1}^2m_\rho^2\over (p_{a_1}\cdot
 p_\rho)^2}|h|^2\right),
 \en
$p$ is the c.m. momentum of $\ov K$ or $a_1$ in the $D$ rest
frame, and $p'$ is the c.m. momentum of the $\rho$ or $\pi$ in the
$a_1$ resonance rest frame.

The resonant three-body decay rate is given by
  \be \label{3body}
 \Gamma(D\to \ov K a_1\to \ov K\rho\pi) &=& {1\over
 8m_D^3}\int^{(m_D-m_K)^2}_{(m_\rho+m_\pi)^2}{dq^2\over 2\pi}\,|M(D\to\ov Ka_1)|^2
 \,|M(a_1\to \rho\pi)|^2   \\
 &\times& {\lambda^{3/2}(m_D^2,q^2,m_K^2)\over 8\pi m_D^2}\,
 {\lambda^{1/2}(q^2,m_\rho^2,m_\pi^2)\over 12\pi q^2}\,
 {1\over (q^2-m_{a_1}^2)^2+(\Gamma_{\rho\pi}(q^2)m_{a_1})^2}, \non
 \en
where $\lambda$ is the usual triangluar function
$\lambda(a,b,c)=a^2+b^2+c^2-2ab-2ac-2bc$, and the ``running" or
``comoving" width $\Gamma_{\rho\pi}(q^2)$ is a function of the
invariant mass squared $m_{\rho\pi}^2=q^2$ of the $\rho\pi$ system
and it has the expression \cite{Pilkuhn}
 \be
 \Gamma_{\rho\pi}(q^2)=\Gamma_{a_1}\,{m_{a_1}\over m_{\rho\pi}}\left({p'(q^2)\over
 p'(m_{a_1}^2)}\right)^3\,{1+R^2p'^2(m_{a_1}^2)\over 1+R^2p'^2(q^2)},
 \en
where $p'(q^2) =
\lambda^{1/2}(q^2,m_\rho^2,m_\pi^2)/(2\sqrt{q^2})$ and we follow
\cite{CLEO1} to take $R$, the ``radius" of the meson, to be
$1.5\,{\rm GeV}^{-1}$. When the resonance width $\Gamma_{a_1}$ is
narrow, the expression of the resonant decay rate can be
simplified by applying the so-called narrow width approximation
 \be
 {1\over (q^2-m_{a_1}^2)^2+m_{a_1}^2\Gamma_{\rho\pi}^2(q^2)}\approx {\pi\over
 m_{a_1}\Gamma_{a_1}}\,\delta(q^2-m_{a_1}^2).
 \en
It is easily seen that this leads to the factorization relation
Eq. (\ref{fact}) for the resonant three-body decay.

Assuming that $|M(D\to \ov Ka_1)|^2$ and $|M(a_1\to\rho\pi)|^2$
are insensitive to the $q^2$ dependence when the resonance is off
its mass shell, these terms will be dropped in the expression of
the parameter $\eta$. We find $\eta=1.07$ and 1.22 for
$\Gamma_{a_1(1260)}=250$ MeV and 600 MeV, respectively. Note that
our results disagree with \cite{Kamal} where the $a_1(1260)$ mass
smearing procedure leads to {\it lower} the rate. The finite width
effect becomes small for $b_1(1235)$, $K_1(1270)$ and $K_1(1400)$
production.

As stressed in \cite{Chengf0}, the finite width effect is most
dramatic when the decay is marginally or even not allowed
kinematically. For example, it is found that $\eta\sim 4.3$ for
$D^0\to f_0(1370)\ov K^0$ for $m_{f_0(1370)}=1370$ MeV and
$\Gamma_{f_0(1370)}=500$ MeV. Evidently, the finite width effect
of $f_0(1370)$ is very crucial for $D^0\to f_0(1370)\ov K^0$.
Recently, the branching ratios of $D^+\to\ov K^{*0}a_1^+(1260)$
and $D_s^+\to\phi a_1^+(1260)$ have been measured by FOCUS
\cite{FOCUS} based on the hypothesis that five-body modes are
dominated by quasi-two-body decays. These modes are not
kinematically allowed if $a_1(1260)$ is very narrow and on its
mass shell. A study of these decays will appear in a forthcoming
publication.

\section{Conclusions}
Cabibbo-allowed charmed meson decays into a pseudoscalar meson and
an axial-vector meson are studied. The charm to axial-vector meson
transition form factors are evaluated in the
Isgur-Scora-Grinstein-Wise quark model. The main conclusions are:

 \begin{enumerate}

 \item The $D\to A$ transition form factor $c_+$ has an opposite
 sign in the ISGW model and its improved
version. It is found that the magnitude of the $D\to\, ^3P_0$ form
factor $V_0$ in the ISGW2 model is three times larger than that in
the ISGW model.

 \item The early predictions of $D^0\to
K^-a_1^+$ and $D^0\to\ov K^0 a_1^+$ rates are too small by a
factor of 5-6 and 2, respectively, when compared with experiment.
The dipole momentum dependence of the form factor for the $D\to K$
transition, which is required by heavy quark symmetry, and the
presence of a sizable long-distance $W$-exchange induced from
final-state rescattering are the two key ingredients for
understanding the data of $D\to \ov Ka_1$. We predict that
$\B(D^+\to \ov K^0a_1^+(1260))=12.1\%$, which is consist with E691
but slightly larger than the Mark III measurement. Experimentally,
it is important to have a refined measurement of this decay mode.

 \item $D\to\ov Kb_1(1235)$ decays are in general suppressed relative to
 $D\to\ov Ka_1(1260)$ owing to the smallness of the decay constant of
 $b_1$. However, the neutral modes $\ov K^0b^0_1(1235)$ and $\ov
 K^0a_1^0(1260)$ are comparable.

 \item The $K_{1A}-K_{1B}$ mixing angle of the strange axial-vector mesons is
extracted  from $\tau\to K_1\nu_\tau$ decays to be $\approx
37^\circ$ or $58^\circ$ with a two-fold ambiguity. This is
consistent with the mixing angles obtained from the experimental
information on masses and the partial rates of $K_1(1270)$ and
$K_1(1400)$. It is found that the positive mixing-angle solutions
are excluded by the study of $D\to K_1(1270)\pi,~K_1(1400)\pi$
decays. An observation of the decay $D^0\to K_1^-(1400)\pi^+$  at
the level of $5\times 10^{-4}$ will rule out $\theta\approx
-37^\circ$ and favor the other solution $\theta\approx -58^\circ$.

 \item Though the decays $D^0\to \ov
K_1^0\pi^0$ are color suppressed, they are comparable to and even
larger than the color-allowed counterparts: $\ov
K_1^0(1270)\pi^0\sim K_1^-(1270)\pi^+$ and $\ov K_1^0(1400)\pi^0>
K_1^-(1400)\pi^+$. It is expected that the neutral modes $D^0\to
\ov K_1^0\pi^0$ have a branching ratio of order $10^{-2}$ and
hence they should be easily accessible by experiment.

 \item The finite width effect of the axial-vector resonance is studied
 It becomes important for $a_1(1260)$ especially when its width
 is near 600 MeV: The $D\to \ov
Ka_1(1260)$ rate is enhanced by a factor of 1.07 and 1.22,
respectively, for $\Gamma_{a_1}=200$ MeV and 600 MeV.

 \end{enumerate}

\vskip 2.5cm \acknowledgments This work was supported in part by
the National Science Council of R.O.C. under Grant No.
NSC91-2112-M-001-038.

\newpage
\centerline{\bf APPENDIX}
\renewcommand{\thesection}{\Alph{section}}
\renewcommand{\theequation}{\thesection\arabic{equation}}
\setcounter{equation}{0} \setcounter{section}{0} \vskip 0.5 cm
\vskip 0.3cm

\section{Form factors in the ISGW model}
Consider the transition $D\to A$, where the axial-vector meson $A$
has the quark content $q_1\bar q_2$ with $\bar q_2$ being the
spectator quark. We begin with the definition \cite{ISGW2}
 \be \label{Fn}
 F_n=\left({\tilde m_A\over \tilde
 m_D}\right)^{1/2}\left({\beta_D\beta_A\over
 \beta_{DA}}\right)^n\left[1+{1\over
 18}h^2(t_m-t)\right]^{-3},
 \en
where
 \be
 h^2={3\over 4m_cm_1}+{3m_2^2\over 2\bar m_D\bar
 m_A\beta_{DA}^2}+{1\over \bar m_D\bar m_A}\left({16\over
 33-2n_f}\right)\ln\left[{\alpha_s(\mu_{\rm QM})\over
 \alpha_s(m_1)}\right],
 \en
$\tilde m$ is the sum of the meson's constituent quarks' masses,
$\bar m$ is the hyperfine-averaged mass, $t_m=(m_D-m_A)^2$ is the
maximum momentum transfer, and
 \be
 \mu_\pm=\left({1\over m_1}\pm {1\over m_c}\right)^{-1},
 \en
with $m_1$ and $m_2$ being the masses of the quarks $q_1$ and
$\bar q_2$, respectively. In Eq. (\ref{Fn}), the values of the
parameters $\beta_D$ and $\beta_A$ are available in \cite{ISGW2}
and $\beta_{DA}^2={1\over 2}(\beta_D^2+\beta_A^2)$.

The form factors defined by Eq. (\ref{DAform}) have the following
expressions in the improved ISGW model:
 \be \label{formISGW2}
 \ell &=& -\tilde m_D\beta_D\left[ {1\over \mu_-}+{m_2\tilde
 m_A(\tilde\omega-1)\over \beta_D^2}\left({5+\tilde\omega\over
 6m_1}-{1\over 2\mu_-}\,{m_2\over \tilde m_A}\,{\beta_D^2\over
 \beta^2_{DA}}\right)\right]F_5^{(\ell)},  \non \\
 c_++c_- &=& -{m_2\tilde m_A\over 2m_1\tilde
 m_D\beta_D}\left(1-{m_1m_2\over 2\tilde
 m_A\mu_-}\,{\beta_D^2\over \beta^2_{DA}}\right)F_5^{(c_++c_-)},
 \non \\
  c_+ -c_- &=& -{m_2\tilde m_A\over 2m_1\tilde
 m_D\beta_D}\left({\tilde\omega+2\over 3}-{m_1m_2\over 2\tilde
 m_A\mu_-}\,{\beta_D^2\over \beta^2_{DA}}\right)F_5^{(c_+-c_-)},
 \\
 r &=& {\tilde m_D\beta_D\over \sqrt{2}}\left[ {1\over
 \mu_+}+{m_2\tilde m_A\over 3m_1\beta^2_D}(\tilde\omega
 -1)^2\right]F_5^{(r)},
 \non \\
 s_++s_- &=& {m_2\over \sqrt{2}\tilde m_D\beta_D}\left( 1-{m_2\over
 m_1}+{m_2\over 2\mu_+}\,{\beta_D^2\over
 \beta^2_{DA}}\right)F_5^{(s_++s_-)}, \non \\
  s_+-s_- &=& {m_2\over \sqrt{2}\tilde m_D\beta_D}\left( {4-\tilde\omega\over 3}-
  {m_1m_2\over 2\tilde m_A\mu_+}\,{\beta_D^2\over
 \beta^2_{DA}}\right)F_5^{(s_+-s_-)}, \non
 \en
where
 \be
 && F_5^{(\ell)}=F_5^{(r)}=F_5\left({\bar m_D\over\tilde
 m_D}\right)^{1/2}\left({\bar m_A\over \tilde m_A}\right)^{1/2},
 \non \\
 && F_5^{(c_++c_-)}=F_5^{(s_++s_-)}=F_5\left({\bar m_D\over\tilde
 m_D}\right)^{-3/2}\left({\bar m_A\over \tilde m_A}\right)^{1/2},
 \non \\
 && F_5^{(c_+-c_-)}=F_5^{(s_+-s_-)}=F_5\left({\bar m_D\over\tilde
 m_D}\right)^{-1/2}\left({\bar m_A\over \tilde m_A}\right)^{-1/2},
 \en
and
 \be
 \tilde \omega-1=\,{t_m-t\over 2\bar m_D\bar m_A}.
 \en

In the original version of the ISGW model \cite{ISGW}, the
function $F_n$ has a different expression in its $(t_m-t)$
dependence:
  \be \label{oldFn}
 F_n=\left({\tilde m_A\over \tilde
 m_D}\right)^{1/2}\left({\beta_D\beta_A\over
 \beta_{DA}}\right)^n{\rm exp}\left\{-{m_2\over 4 \tilde m_D\tilde
 m_A}\,{t_m-t\over \kappa^2\beta_{DA}^2}\right\},
 \en
where $\kappa=0.7$ is the relativistic correction factor. The form
factors are then given by
 \be
 \ell &=& -\tilde m_D\beta_D\left[ {1\over \mu_-}+{m_2\over 2\tilde m_D}\,
 {t_m-t\over \kappa^2\beta_D^2}\left({1\over m_1}-{1\over 2\mu_-}\,
 {m_2\over \tilde m_A}\,{\beta_D^2\over
 \beta^2_{DA}}\right)\right]F_5,  \non \\
  c_+&=& {m_2m_c\over 4\tilde m_D\beta_D\mu_-}\left(1-{m_1m_2\over 2\tilde
 m_A\mu_-}\,{\beta_D^2\over \beta^2_{DA}}\right)F_5,
 \non \\
 s_+ &=& {m_2\over \sqrt{2}\tilde m_D\beta_D}\left( 1+{m_c\over
 2\mu_-}-{m_1m_2m_c\over 4\mu_+\mu_-\tilde m_A}\,{\beta_D^2\over
 \beta^2_{DA}}\right)F_5.
 \en
It is clear that the form factor $c_+$ has an opposite sign in the
ISGW and ISGW2 models. Note that the expressions in Eq.
(\ref{formISGW2}) in the ISGW2 model allow one to determine the
form factors $c_-$ and $s_-$, which vanish in the ISGW model.


\end{document}